\documentclass[prb,preprintnumbers,superscriptaddress,shownopacs,twocolumn,balancelastpage,floatfix]{revtex4}

\usepackage{amsmath,amssymb,amsfonts}
\usepackage{bm,graphicx,color}

\newcommand{\TR}{\text{Tr}}

\newcommand{\Vk}{{\bm k}}

\newcommand{\expval}[1]{\langle{#1}\rangle}

\newcommand{\CC}{\mathcal{C}}

\newcommand{\convz}{\ast}

\newcommand{\les}{<}

\newcommand{\one}{{\bm I}}

\newcommand{\teff}{T_\text{eff}}
\newcommand{\Etot}{\mathcal{E}}
\newcommand{\tterm}{\tau_\text{th}}

\begin{document}

  \title{Thermalization of a pump-excited Mott insulator}

  \author{Martin Eckstein}
  \author{Philipp Werner}
  
  \affiliation{Theoretical Physics, ETH Zurich, 8093 Zurich, Switzerland}  
  
  \date{\today}

\begin{abstract}
We use nonequilibrium dynamical mean-field theory in combination with  a recently implemented
strong-coupling impurity solver to investigate the relaxation of a  Mott insulator after a laser
excitation with frequency comparable to the Hubbard gap. The time  evolution of the double
occupancy exhibits a crossover from a strongly damped transient at  short times towards an exponential
thermalization at long times. In the limit of strong interactions, the  thermalization time is
consistent with the exponentially small decay rate for artificially  created doublons, which was measured
in ultracold atomic gases. When the interaction is comparable to the bandwidth, on  the other hand, the double
occupancy thermalizes within a few times the inverse bandwidth along a  rapid thermalization path
in which the exponential tail is absent. Similar behavior can be  observed in time-resolved
photoemission spectroscopy. Our results show that a simple quasi-equilibrium description of the
electronic state breaks down for pump-excited Mott insulators  characterized by strong interactions.
\end{abstract}

\maketitle

\section{Introduction}

State-of-the-art pump-probe experiments use ultrashort light pulses to control and probe the 
complex interplay between electronic, magnetic and lattice degrees of freedom in strongly 
correlated materials.\cite{Iwai2003a,Perfetti2006a,Rini2007,Wall2011,Dean2011,Ichikawa2011}  
The Mott metal-insulator transition, which is the paradigm for phase transitions in correlated 
electron systems, has been investigated in several time-resolved experiments. For example, 
a photo-induced metal-insulator transition and the ultrafast recovery of the Mott gap was 
observed in a charge-transfer insulator by using reflectivity measurements,\cite{Iwai2003a} 
and in 1T-TaS$_2$ using time-resolved photoemission spectroscopy.\cite{Perfetti2006a} 
Recently, Wall {\em et al.} succeeded in measuring electronic dynamics in an organic 
charge-transfer insulator on the timescale of the inverse bandwidth $\hbar/W$.\cite{Wall2011}

Photo-excited phases in materials with several competing interactions have been reached 
along nonthermal pathways,\cite{Rini2007} and their properties can be quite 
different from those of any known equilibrium state of the same system.\cite{Ichikawa2011,
Dean2011} On the other hand, in many experiments the electronic state has been 
interpreted in terms of the two-temperature model,\cite{Allen1987a} whose basic 
assumption is that electrons thermalize rapidly (on the timescale of the inverse 
bandwidth) and thereafter can be assigned a temperature which is different from the lattice temperature.
Deviations from the two-temperature behavior have been observed long ago in noble 
metals, where the scattering between quasiparticles is weak and thermalization times can 
range up to a few hundred femtoseconds.\cite{Fann1992a}  In strongly interacting 
quantum systems, thermalization is theoretically not well understood, and it is thus 
unclear whether a homogeneous electron system can be stuck in a nonthermal 
state on timescales much longer than $\hbar/W$. 

Nonequilibrium experiments on ultracold atomic gases\cite{Greiner2002b,Kinoshita2006} have 
recently stimulated considerable theoretical effort to resolve some of the questions related to 
the thermalization of isolated quantum-many body systems.\cite{quenchreview} In several 
models\cite{Kollath2007a,Barmettler2009a,Eckstein2009a,Eckstein2010a} it was found that 
the relaxation after a sudden quench of the Hamiltonian sensitively depends on the interaction 
parameter. In the fermionic Hubbard model, e.g., rapid thermalization of the momentum 
distribution (within  $t \sim \hbar/W$) occurs after quenches from the noninteracting ground state 
to a very narrow interaction regime at intermediate coupling ($U \approx 0.8W$ for the semi-elliptic 
density of states),\cite{Eckstein2009a} while after quenches to weak and strong coupling one 
observes the formation of a quasistationary nonthermal state which persists for $t \gg \hbar/W$.%
\cite{Moeckel2008a,Eckstein2009a,Eckstein2010a} The latter phenomenon, which is called prethermalization, 
can be related to the non-ergodic behavior of integrable models,\cite{Rigol2007a} because in 
both cases the dynamics is constrained by either exact or approximate constants of 
motion.\cite{Kollar2011} 

 In this paper we turn from interaction quenches to the pump-probe setup and study the relaxation of a 
 Mott insulator after a short laser excitation. The analysis is performed for the Hubbard model,
 \begin{equation}
 \label{hubbard}
 H= \sum_{ij,\sigma=\uparrow,\downarrow} \!\!t_{ij}\, c_{i\sigma}^\dagger c_{j\sigma}
 + U
  \sum_{i}
  \big(n_{i\uparrow}\!-\!\tfrac12\big)
  \big(n_{i\downarrow}\!-\!\tfrac12\big),  
 \end{equation}
 which describes fermionic particles that can hop between the sites of a crystal 
 lattice (with hopping amplitude $t_{ij}$) and interact with each other through a local 
 Coulomb repulsion $U$. In the phase diagram of the Hubbard model there is 
 a metal-insulator transition at low temperatures and half-filling ($n_{\uparrow}
 =n_\downarrow=\tfrac12$). For the present investigation we focus on higher 
 temperatures, where this phase transition turns into a smooth crossover, and 
 study the relaxation after a short pump-pulse that excites carriers over the Mott gap.
 Although the equilibrium states before the excitation are then continuously connected, 
 the response to the pulse turns out to be strongly dependent on the value of $U$. 
 
 For all values of $U$, the system approaches a thermal state, which is at elevated 
 temperature because the energy is increased during the pump and conserved thereafter.
 However,  just as for the interaction quench\cite{Eckstein2009a} the thermalization time 
 $\tterm$ depends sensitively on $U$. Already for moderately strong interactions, where 
 the Hubbard gap is still comparable to the bandwidth, $\tterm$ is much larger than $\hbar/W$, 
 which puts a question mark behind the applicability of a simple two-temperature model 
 in this case. At intermediate coupling, on the other hand, thermalization indeed happens 
 on the timescale of the inverse bandwidth, but it then follows a qualitatively different path 
 which may be referred to as ``rapid thermalization''.
 
 To study the dynamics of the Hubbard model, we use the dynamical mean-field 
 theory,\cite{Georges1996}  which provides a mapping of the lattice model onto a single 
 impurity model and is exact in the limit of infinite dimensions.\cite{Metzner1989a} The 
 reformulation of DMFT within the Keldysh framework\cite{Schmidt2002,Freericks2006a} allows to apply this approach to 
 nonequilibrium problems in which a system is initially prepared in a thermal equilibrium state and 
 subsequently exposed to some perturbation. The biggest challenge for the method is the solution 
 of a quantum impurity model under quite general nonequilibrium conditions. A generalization of 
 the auxiliary-field weak-coupling continuous-time quantum Monte Carlo method\cite{Werner2009a}
 has been used successfully to study the short-time dynamics of the Hubbard model,\cite{Eckstein2009a,
 Eckstein2010a,Tsuji2010,Eurich2010} but the dynamical sign problem prevents an investigation 
 of the long-time behavior, in particular for large values of $U$ which are of interest for the 
 current investigation of Mott-insulating states. Possibilities to avoid the difficulties of real-time
 Monte Carlo methods  include the superperturbation theory,\cite{Jung2011} and perturbative
 approaches.\cite{Eckstein2010a,Eckstein2010b} For the present study we use the self-consistent
 strong-coupling expansion,\cite{Eckstein2010b} which is a generalization of the non-crossing
 approximation\cite{NCA} (NCA) to higher orders and to the Keldysh contour.
 It works particularly well in the Mott insulating phase, and the  comparison of results 
 from different orders (up to third order)  provides at least some estimate of the error.

The outline of this paper is as follows: In Section \ref{sec::model} we give some details of the 
DMFT framework used here, in particular the solution of the lattice Dyson equation in real
(Keldysh) time. We then present results for the relaxation of the double occupancy after a
pump-excitation of a Mott insulator (Sec.~\ref{sec::docc}) and identify the corresponding 
signatures in time-resolved photoemission spectroscopy (Sect.~\ref{sec::spec}). A 
conclusion is given at the end (Sec.~\ref{sec::conclusion}).

\section{The Model}
\label{sec::model}

 \subsection{Hubbard Model in an external electric field}

  In this paper  we investigate the Hubbard model with an additional spacially 
  homogeneous, but time-dependent electric field,
  \begin{equation}
  \label{efield}
  \bm{E}(t) = \bm{\hat n}\,E_0 \sin(\Omega t) \phi(t),
  \end{equation}
  corresponding to a few-cycle pump-pulse with amplitude $E_0$, polarization $\bm{\hat n}$, 
  frequency $\Omega$, and a pulse envelope $\phi(t)$. (The assumption of spacial homogeneity 
  is appropriate for optical frequencies with wavelength much larger than the lattice  spacing.)
  Within a gauge without scalar potential, $\bm{E}$ is determined by the vector potential 
  through the relation $\bm{E}=-\frac{1}{c} \partial_t \bm{A}$, and the latter is  incorporated into 
  the Hamiltonian (\ref{hubbard}) via a Peierls substitution in the hopping amplitudes $t_{ij}$. This 
  leads to a shift of the band energies $\epsilon_\Vk$, i.e., the hopping part of Eq.~(\ref{hubbard}) 
  becomes time-dependent,
  \begin{equation}
  \label{hubbard-hopp}
  H(t) = \sum_{\Vk\sigma}  h_\Vk(t) n_{\Vk\sigma}
  +
    U
    \sum_{i}
    \big(n_{i\uparrow}\!-\!\tfrac12\big)
    \big(n_{i\downarrow}\!-\!\tfrac12\big),  
  \end{equation}
  with $h_\Vk(t) = \epsilon_{\Vk- \frac{e}{\hbar c}\bm{A}(t)}$. In the following we choose a hypercubic lattice in the 
  limit of infinite dimensions with a Gaussian density of states 
  \begin{equation}
  \label{hcdos}
  \rho(\epsilon) = \frac{1}{\sqrt{\pi}W}\exp\big(-\frac{\epsilon^2}{W^{2}}\big). 
  \end{equation}
  The unit for energy is given by the variance  $W$ of the density of states, which will be loosely referred 
  to as the bandwidth, time is measured in units of $\hbar/W$, and the unit of the electric field is given by 
  $W/ea$, where $-e$ is the electronic charge and $a$ is the lattice spacing. In these units, the critical 
  end-point of the first-order Mott transition line in the phase diagram of the paramagnetic half-filled 
  Hubbard model, which is obtained quite accurately at second order in the self-consistent
  hybridization expansion,\cite{Eckstein2010b}  is located at $U \approx 3.1$ and $T\approx 0.02$.
  
  The DMFT equations for this setup have been discussed in detail in Ref.~\onlinecite{Freericks2006a}
  for a polarization $\bm{\hat n}=(1,1,\ldots ,1)^t$ along the body diagonal of the cubic unit cell, and they
  will not be repeated in detail. Because $\bm{A}(t)$ enters in the effective impurity action only indirectly 
  via the hybridization function, we can apply the self-consistent hybridization expansion as an impurity 
  solver without modifications.\cite{Eckstein2010b} If not stated otherwise we use the second order of that approximation  
  as an impurity solver (one-crossing approximation, OCA), which is reliable in the insulating phase 
  and in the crossover regime.\cite{Eckstein2010b}
 
  \subsection{Solution of the DMFT self-consistency}
 
  Within the DMFT framework, a repeated solution of the lattice Dyson equation and the Dyson equation of 
  the impurity model is required (for details, see, e.g., Refs.~\onlinecite{Freericks2006a} and \onlinecite{Eckstein2010a}).
  These equations  are in essence integro-differential equations on the Keldysh 
  contour $\CC$ which runs from $0$ to time $t_\text{max}$ (i.e., the largest time of interest) along the real axis, 
  back to $0$, and finally to $-i\beta$ along the imaginary time axis.\cite{keldyshintro} Since there are in principle
  several ways to rephrase and solve these equations, we use the 
  remaining paragraphs of this section to describe a scheme which we found to be reliable. The same scheme 
  has been used previously to study the dielectric breakdown of the Mott insulator.\cite{Eckstein2010c}
  
  The following approach is mainly designed to satisfy two properties: 
  (i) It is easy to implement a higher order scheme in the timestep $\Delta t$ to reduce the 
  discretization error. We found that this considerably 
  increases the accuracy for a given number of timesteps $N$ on $\CC$, which in turn strongly reduces
  the required storage [$\mathcal{O}(N^2)$] and computational effort [$\mathcal{O}(N^3)$ for NCA, 
  $\mathcal{O}(N^4)$ for OCA]. (ii) The approach exactly satisfies the causal structure of the equations, 
  i.e., the solution at a given time is computed using only information from previous times. This makes the 
  solution of the nonequilibrium DMFT equations equivalent to a time-propagation that starts from the initial 
  equilibrium DMFT solution, rather than an iteration of the equations on the full contour $\CC$.
   
  To design such a scheme we rewrite the integro-differential equations on $\CC$ explicitly for the 
  imaginary-time, real-time, and mixed imaginary-time/real-time components of the two-time 
  propagators $G(t,t')$. This step has been described in detail in Ref.~\onlinecite{Eckstein2010a}. 
  The resulting equations are integral equations of Volterra type, whose solution can be easily 
  designed in a  way to satisfy the requirements (i) and (ii) above. We found that two types of integral 
  equations are particularly well-behaved (both are to be solved for $\bm X$): 
  The integral equation
  \begin{equation}
  \label{vie2}
  (\one - {\bm A}) \convz {\bm X}  = {\bm B},	
  \end{equation}
  and the integro-differential equation (basically a Dyson equation)
  \begin{equation}
  \label{dyson}
  [i\partial_t - {\bm h} - {\bm A}] \convz {\bm X}  = \one,
  \end{equation}
  where ${\bm h}(t,t')=\delta_\CC(t,t')h(t)$ is a function of the physical time $t$ only. 
  Here and in the following, bold quantities $\bm X(t,t')$ denote two-time functions  with 
  arguments on $\CC$, $[\bm A \convz \bm B](t,t') = \int_\CC d\bar t\,
  \bm A(t,\bar t) \bm B(\bar t,t')$ is the contour convolution, and 
  $\one(t,t')=\delta_\CC(t,t')$ is the delta function on $\CC$.
  When written as explicit integral equations for the real and 
  imaginary-time components, Eq.~(\ref{vie2}) leads to Volterra
  equations of the second kind. In contrast, the slightly modified
  equation ${\bm A} \convz {\bm X} = {\bm B}$ leads to Volterra equations 
  of the first kind, which tend to be unstable for high-order algorithms.
  In the remainder of this section we show how to rewrite the DMFT 
  selfconsistency equations (appropriate for the NCA-type impurity solver) 
  only in terms  of equations of type (\ref{vie2}).
  
  Nonequilibrium DMFT corresponds to a self-consistent solution of the 
  following set of equations: 
 %
  (i) Solution of the impurity problem, which in the case of the strong-coupling 
  expansion requires an evaluation of the diagrammatic expression for the local Greenfunction in 
  terms of the  hybridization function $\bm \Lambda$,\cite{Eckstein2010b}
  \begin{equation}
  {\bm G} = {\bm G}_{NCA,\ldots}[\bm \Lambda],
  \end{equation}
  (ii), the Dyson equation of the impurity model
  \begin{equation}
  \label{impdyson}
  [i\partial_t  + \mu - {\bm \Sigma} - {\bm \Lambda}] \convz {\bm G} = \one,
  \end{equation}
  (iii), the lattice Dyson equation with the time-dependent band-energies $h_\Vk(t)$ from Eq.~(\ref{hubbard-hopp})
   \begin{equation}
   \label{lattdyson}
  [i\partial_t  + \mu - {\bm \Sigma} - {\bm h}_\Vk] \convz {\bm G}_\Vk = \one,
  \end{equation}
  and (iv), the momentum sum
   \begin{equation}
   \label{ksum}
  \sum_\Vk {\bm G}_\Vk = {\bm G}.
  \end{equation}
  To reformulate this set of equations only in terms of equations of type (\ref{vie2})
  we introduce the quantity $\bm Z=[i\partial_t + \mu - {\bm \Sigma}  ]^{-1}$,
  conjugate to the self-energy, and the momentum sums
   $\bm G_1 = \sum_\Vk \bm h_\Vk {\bm G}_\Vk $,
   $\bm G_1^\dagger = \sum_\Vk {\bm G}_\Vk \bm h_\Vk $, and 
   $\bm G_2 = \sum_\Vk \bm h_\Vk {\bm G}_\Vk \bm h_\Vk$
   in addition to Eq.~(\ref{ksum}). By summing Eq.~(\ref{lattdyson}) over $\Vk$
   (using $\sum_\Vk h_{\Vk} =0$) and comparing with Eq.~(\ref{impdyson}) one 
   can show that  $\bm G_1= \bm \Lambda \convz \bm G$, $\bm G_1^\dagger=   
   \bm G  \convz   \bm \Lambda$, and that Eqs.~(\ref{impdyson})-(\ref{ksum}) 
   are equivalent to
   \begin{align}
   \label{k-1}
   &
   [\one + {\bm G}_1^\dagger] \convz {\bm Z} = {\bm G}, 
   \\
   \label{k-2}
   &
   [\one - {\bm F}_\Vk] \convz {\bm G}_\Vk = {\bm Z},
   \\
   \label{k-3}
   &
   [\one + \bm G_1] \convz \bm \Lambda = \bm G_2,
   \end{align}
   where ${\bm F}_\Vk(t,t') = {\bm Z}(t,t')h_\Vk(t')$.
   In practice, we determine $\bm G$ from $\bm \Lambda$
   via the self-consistent hybridization expansion, perform the 
   convolution  $\bm G \convz \bm \Lambda$ to obtain
   an estimate for $\bm G_1^\dagger$, then solve
   Eq.~(\ref{k-1}) for $\bm Z$, Eq.~(\ref{k-2}) for ${\bm G}_\Vk$
   on a given set of $\Vk$-points, perform the momentum sums
   to obtain $\bm G$, $\bm G_1$, $\bm G_1^\dagger$, and $\bm G_2$, and 
   finally solve Eq.~(\ref{k-3}) for $\bm \Lambda$. This cycle is 
   performed consecutively for each timestep (starting from an 
   extrapolation of $\Lambda$ from the previous timestep), which 
   usually does not require more than two or three iterations to 
   reach a converged solution.

 \section{Thermalization of the pump-exited Mott insulator} 
  \label{sec::docc}
  
  \begin{figure}
  \centerline{\includegraphics[clip=true,width=0.95\columnwidth]{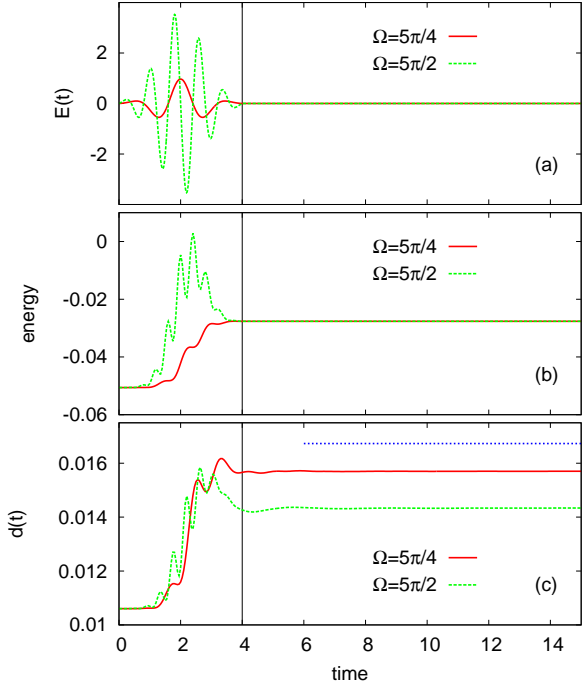}}
  \caption{
  Pump-excitation of a Mott insulator at $\beta=5$, $U=5$.
  a) The electric field used for the data of this plot. The two frequencies $\Omega=5\pi/4
  \approx3.927$ and $\Omega=5\pi/2\approx7.854$ are comparable to the Hubbard gap 
  and larger than the Hubbard gap, respectively, and the amplitudes $E_0=0.9679$ ($\Omega
  =5\pi/4$) and  $E_0=3.6701$ ($\Omega=5\pi/2$) are chosen such that $\teff=0.5$,
  i.e., the energy of the pump-excited system for $t>4$ is the same as in a system in 
  thermal equilibrium at temperature $T=\teff$.
  b) Time-dependence of the total energy $\Etot$. c) Time dependence of the double 
  occupancy $d(t)$. The horizontal dotted line indicates the double occupancy $d(\teff)$ 
  in an equilibrium state at temperature $T=\teff$.}
  \label{fig-pulse}
  \end{figure}  	
 
  In this section we analyze the dynamics of Mott-insulating and crossover states in the half-filled 
  Hubbard model after pump-pulses with frequencies of the order of the gap.  The system is initially 
  prepared in an equilibrium state at given temperature $T=1/\beta$, and excited by a few-cycle 
  electrical field pulse [Eq.~(\ref{efield})] that has a Gaussian pulse envelope $\phi(t) = \exp[-(t-2)^2]$ 
  and is restricted to time $0\le t \le 4$ (Fig.~\ref{fig-pulse}a).  Figure \ref{fig-pulse}b shows the 
  typical time dependence of the total energy $\Etot(t) = \expval{H(t)}$ during and after the 
  pump-excitation for an initially Mott-insulating state ($U=5$, $\beta=5$). During the pump, $\Etot(t)$ 
  increases with time due to the absorption of light, while it is constant or $t>4$ because the system 
  is not coupled to external thermal reservoirs. Note that energy conservation comes out correct only 
  if all approximations that are made for the solution of the impurity model are conserving in the sense 
  of Kadanoff and Baym, which is true for the self-consistent hybridization expansion. More precisely, 
  energy conservation implies the relation $ \dot \Etot(t) = \bm{E}(t)\expval{\bm{j}(t)}$, which can serve 
  as a first check for the numerics ($\bm{j} = e\sum_{\Vk\sigma} n_{\Vk\sigma} \bm{v}_\Vk$ is the current 
  operator, with band velocities $\bm{v}_\Vk =\hbar^{-1}\partial_\Vk \epsilon_{\Vk-\frac{e}{\hbar c}\bm{A}}$).

  It is the basic assumption of the two-temperature model that due to the electron-electron interaction 
  electrons in a solid thermalize so fast that the lattice dynamics is not important for that process. In 
  the present case, thermalization implies that the time-evolved state has the same properties as the 
  uniquely defined thermal equilibrium state whose temperature is such that its energy equals the energy 
  of the pump excited system, $\Etot(t>4)=\TR[e^{-H/\teff} H]/Z$. The latter condition defines an effective 
  temperature $\teff(\Etot)$, and we test for thermalization by comparing time-dependent expectation 
  values of observables $\mathcal{O}(t)$ to thermal expectation values $\mathcal{O}(\teff)=\TR[e^{-H/\teff} 
  \mathcal{O}]/Z$ which are obtained from an independent equilibrium DMFT calculation. To compute 
  the thermal energy $\Etot(T)=\TR[e^{-H/T} H]/Z$ which is needed for the determination of $\teff$, 
  we use the same impurity solver as in the nonequilibrium calculation, i.e., the same order of the 
  self-consistent hybridization expansion.
  
  In Fig.~\ref{fig-pulse}c we compare thermal and time-dependent expectation values
  of the double occupancy $d(t) = \expval{n_{i\uparrow}n_{i\downarrow}}$, which is a 
  purely local quantity and can be measured directly in the impurity model. Apparently, 
  a large part of the excitation energy goes into the creation of doublon-holon pairs, 
  such that $d(t)$ rises by more than $30\%$ during the pump relative to its small 
  value in the initial Mott-insulating state. Subsequent to the pump, $d(t)$ rapidly settles
  at a value $d_\text{stat} \neq d(\teff)$, and further relaxation towards the thermal value 
  is hardly visible on the scale of Fig.~\ref{fig-pulse}. This behavior is similar to that of 
  the Hubbard model after an interaction quench.\cite{Eckstein2009a} However, a sudden 
  change of the interaction causes pronounced $2\pi/U$-periodic oscillations which have 
  been used to characterize the short time behavior in Ref.~\onlinecite{Eckstein2009a}, while these oscillations are very weak in 
  the present case, where the pump-excitation itself takes already longer than the period 
  $2\pi/U$.

   \begin{figure}
   \centerline{\includegraphics[width=0.99\columnwidth]{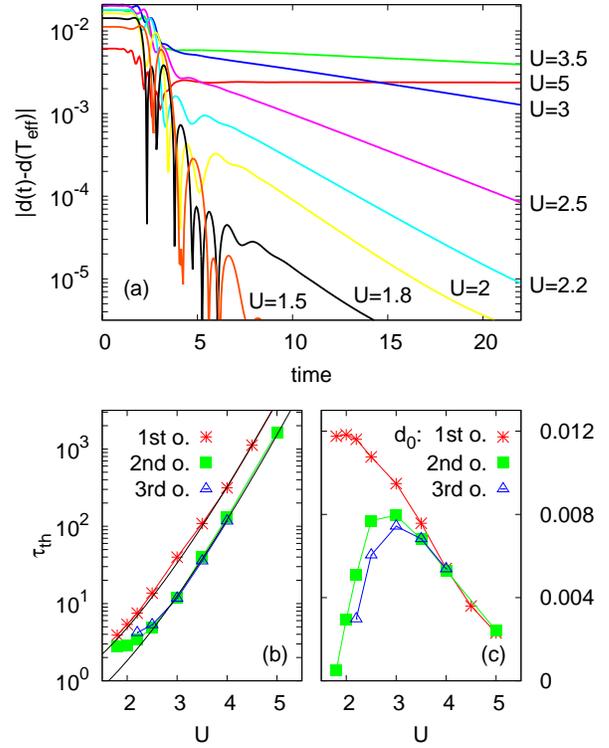}}
   \caption{
  Relaxation of the double occupancy $d(t)$ after an excitation with frequency
  $\Omega=U\pi/2$ ($\beta=5$). Upper panel: (a) The difference $|d(t)-d(\teff)|$,
  obtained from the OCA impurity solver. The pump-amplitudes
   $E_0$ are given by 3.6701 ($U=5$), 2.4397 ($U=3.5$), 2.1325 ($U=3$), 1.9123 ($U=2.5$), 
  1.8198 ($U=2.2$), 1.7761 ($U=2$), 1.6632 ($U=1.8$), 1.2765 ($U=1.5$), such that $\teff=0.5$
  in all cases. Lower panels: Parameters $\tterm$ (b) and $d_0$ (c), obtained from fitting 
  the curves in a) for $t>12$ with Eq.~(\ref{linfit}) (square symbols). Stars and triangles 
  show the result of the analogous investigation using NCA and the 3rd order hybridization 
  expansion, respectively. The time-evolution in 3rd order was restricted to $t \le 15$.
  Black solid lines in b) correspond to the exponential law\cite{Sensarma2010a}
  $\tterm= \tau_0 \exp(\alpha U \log U )$, with $\alpha=1.01$ and 
  $\tau_0=0.45\, (1.2)$ for OCA (NCA).}
  \label{fig-doccrelax}
  \end{figure} 
  
  The long-lived nonthermal value of $d(t)$ implies that doubly occupied and empty 
  sites do not rapidly recombine in the Hubbard model at large Coulomb repulsion.\cite{Rosch2008,Sensarma2010a}
  In fact, in experiments with ultra-cold atoms it was found that for $U \gg W$ an artificially 
  increased double occupancy relaxes only on the exponentially long timescale $
  \tau \propto  \exp(\alpha \log(U/W) U/W) $,\cite{Sensarma2010a} in agreement with theoretical
  predictions for the lifetime of a single doublon in a background of singly occupied 
  sites and holes. The reason for this is that multi-particle scattering events are required to 
  transform a large quantum of energy ($U$) into many kinetic energy excitations of a typical amount $W$, 
  similar to the case of a semiconductor where electrons and holes cannot effectively 
  recombine if the main relaxation channel is the emission of phonons with a much lower 
  energy than the electronic band gap. The nonequilibrium DMFT approach now allows 
  us to investigate how this decay rate of a single doublon is related to the thermalization 
  of a ``cloud'' of many excitations, and how the relaxation is modified when $U$ and $W$ 
  are comparable in magnitude.

  To answer this question we plot the difference $|d(t)-d(\teff)|$ for a series of initial states 
  at various values of the Coulomb interaction and temperature $T=0.2$ (Fig.~\ref{fig-doccrelax}a). 
  The pump frequency is $\Omega=\pi U/2$, and the pump amplitude $E_0$ is tuned such that the 
  effective temperature after the pump is given by $\teff=0.5$. The time-evolution $d(t)$ 
  is initially dominated by a damped transient, 
  which subsequently turns into a smooth exponential relaxation towards the thermal 
  value $d(\teff)$. The analysis of this exponential tail offers a rigorous way to 
  separate the rapid short-time relaxation from the slower long-time thermalization: From 
  a linear fit to the curves in Fig.~\ref{fig-doccrelax}a (for $t>12$), i.e. an ansatz
  \begin{equation}
  \label{linfit}
  d(t) \sim d(\teff) + d_{0} \exp(-t/\tterm),
  \end{equation}
  we extract both the long-time thermalization rate $1/\tterm$, 
  and a value $d_{0}$ which measures the ``residue'' of the initial rapid relaxation process. 
  In the limit $U \gg W$, $d_0$ becomes identical to the apparent nonthermal stationary double 
  occupancy $d_\text{stat}$ introduced in the discussion of Fig.~\ref{fig-pulse}. The
  numerical effort required to resolve the small difference $d(t)-d(\teff)$ restricts the accessible 
  times, such that  a relaxation over more than one order of 
  magnitude can be observed only for a limited range of interactions ($U=2, 2.2, 2.5$ in 
  Fig.~\ref{fig-doccrelax}). Nevertheless, the behavior of $\tterm$ and $d_0$ as a function 
  of $U$ which is obtained on the basis of our present results already draws an interesting 
  picture for the relaxation of the Hubbard model at strong and intermediate coupling, which 
  we will discuss in the following two paragraphs.
  
  For $U \gg W$, where the total number of excited doublons is small, we find that the 
  dependence of the thermalization rate $\tterm$ on $U$ is indeed consistent with the exponential 
  law\cite{Sensarma2010a} $\tterm = \tau_0 \exp(\alpha U\log U)$ that was predicted for the decay 
  of a single doublon (Fig.~\ref{fig-doccrelax}b). Up to numerical accuracy,  NCA, OCA, and the 
  third order of the hybridization expansion give the same large-$U$ asymptotics for $\tterm$ 
  ($\alpha\approx 1$) and differ only by the prefactor $\tau_0$. Due to the exponential 
  dependence on $U$, the thermalization rate is much larger than the inverse bandwidth 
  already for moderately strong interactions.

  When $U$ and $W$ are similar in magnitude, the relaxation changes in
  two ways: (i) the rate $\tterm$ significantly decreases to values of the order
  of a few inverse $W$ and starts to deviate from the large-$U$ asymptotics, 
  and (ii), the residue $d_0$ extrapolates to zero at a finite 
  value of the interaction. While both facts eventually imply thermalization 
  on a timescale given by the inverse bandwidth, they are conceptually quite different. 
  The vanishing of the residue $d_0$ implies the absence of the long-time exponential
  relaxation tail, such that the system follows a qualitatively different path to the thermal 
  state. In fact, the vanishing of the rapid relaxation residue $d_0$ provides a rigorous 
  way to define a point of ``rapid thermalization'', as opposed to a regular, two step 
  relaxation. For $U<1.5$ thermalization seems to become slower again, in agreement 
  with similar findings for the interaction quench to the weak-coupling regime,\cite{Moeckel2008a,
  Eckstein2009a} but due to the limited numerical accuracy we cannot fit an 
  exponential tail to the curves for $U<1.8$ and study this regime systematically.
  That the system approaches a rapid thermalization point in the 
  current setup may come as a surprise, because neither the sequence of initial states 
  at $T=0.2$, nor that of the the final states at $T=0.5$ crosses a known 
  phase-transition of the half-filled Hubbard model. The phenomenon 
  is clearly beyond a simple quasi-equilibrium explanation, and it is an 
  interesting question whether it is related to the ``rapid thermalization'' at the dynamical
  transition in the Hubbard model after an interaction quench,\cite{Eckstein2009a}
  and whether the latter can be characterized in a similar way.

  \section{Time-resolved photoemission spectroscopy} 
  \label{sec::spec}

  \begin{figure}
  \centerline{\includegraphics[width=0.99\columnwidth]{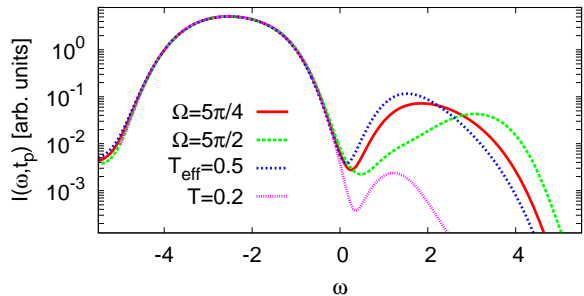}}
  \caption{Time-resolved photoemission spectrum [Gaussian probe, $\delta=2$, $t_p=10$]  
  of a pump-excited system ($U=5$, $\beta=5$), using the pump parameters of Fig.~\ref{fig-pulse}. The
  blue dotted line marked $\teff=0.5$ is the photoemission spectrum of a system at temperature 
  $T=0.5$, for the same Gaussian probe pulse as for the other two curves.}
  \label{fig-spec1}
  \end{figure}  	

  The results of the previous section raise the question how well the different thermalization behaviors 
  at strong and intermediate coupling can be distinguished by means of spectroscopic methods. In this 
  section we will investigate this question for the case of time-resolved photoemission spectroscopy.
  The signal $I(\omega,t_p)$ in this case is simply defined as the probability for an electron to get 
  emitted into an outgoing state with kinetic energy $\hbar \Vk^2/2m = \hbar \omega_\text{probe} - \omega -\Phi$  in response
  to a probe-pulse with frequency $\omega_\text{probe}$ at the probe-time $t_p$ ($\Phi$ is the unknown
  work-function). Similar to equilibrium, the photoemission spectrum can be related to the single-particle 
  Greenfunction $G^\les(t,t')  = i\expval{c_{j\sigma}^\dagger(t')c_{j\sigma}(t)} $ on a lattice 
  site $j$,\cite{Freericks2009a}
  \begin{equation}
  \label{pes}
  I(\omega,t_p)
  =
  -i \!\int \! \! dt \,dt'\, 
  S(t) S(t')
  e^{i\omega(t'-t)}
  G^\les(t_p+t,t_p+t'),
  \end{equation}
  where $S(t)$ is the envelope of the probe pulse. For the following investigation we use a Gaussian probe 
  $S(t)=\exp(-t^2/2\delta^2) \Theta(3\delta - |t| )$, where $\delta$ is the duration. We cut off the exponential 
  tails of the probe for $|t|>t_c=3\delta$, such that Eq.~(\ref{pes}) can be evaluated for probe times 
  $t_c \le t_p \le t_\text{max}-t_c$ if the Greenfunction is known for $0 \le t,t' \le  t_\text{max}$.
  
  For the derivation of Eq.~(\ref{pes}) one has to disregard matrix element effects and use the sudden 
  approximation, which neglects the interaction between electrons in the outgoing states and in the 
  solid.\cite{Freericks2009a} These approximations are the same that are commonly made in order to compare photoemission spectra in 
  equilibrium to the product of the spectral function $A(\omega)$ and the Fermi
  function $f(\omega)$. In fact, for an equilibrium state, $G^\les(t,t')$ is just given by the Fourier transform
  $G^\les(t,t') = i \int d\omega A(\omega)f(\omega)e^{i\omega(t'-t)}$, such that Eq.~(\ref{pes}) is reduced
  to a convolution of the product $A(\omega)f(\omega)$ with the  power density $|\tilde S(\omega)|^2= 
  |\int dt S(t)e^{i\omega t}  |^2$ of the probe envelope $S(t)$, i.e., $I(\omega) = \int d\omega' |\tilde S (\omega')|^2 
  A(\omega-\omega')f(\omega-\omega')$. Hence $A(\omega)$ can in principle be measured 
  with arbitrary accuracy by choosing a sufficiently long probe-pulse. On the other hand, for the 
  nonequilibrium case the intrinsic frequency-time uncertainty in Eq.~(\ref{pes}) makes it 
  impossible to measure (or even define) transport of spectral weight  on the timescale of the 
  inverse bandwidth.\cite{Eckstein2008c} In the present case, however, we will show that
  time-resolved photoemission can be used to identify the thermalization signatures 
  that have been discussed in the previous section.
   
   In Fig.~\ref{fig-spec1} we reconsider the two pump-experiments of Fig.~\ref{fig-pulse} 
   on a Mott-insulating initial state ($U=5$, $\beta=5$), with different pump-frequencies 
   but an equal amount of  absorbed energy ($\Omega=\pi U/4$ and $\Omega=\pi U/2$, 
   $\teff=0.5$). The photoemission spectrum $I(\omega,t_p)$ [Eq.~\ref{pes})] is broadened 
   due to a relatively short pulse duration $\delta=2$, but the Hubbard bands and a gap can
   still clearly be distinguished. Similar to the slow relaxation of the double occupancy 
   at $U=5$, $I(\omega,t_p)$ becomes almost independent of $t_p$ after the decay of the 
   initial transient for all times accessible with the OCA impurity solver. In the figure, we compare 
   the spectra of the pump-excited system at $t_p=10$ to equilibrium spectra both at the initial
   temperature ($T=0.2$) and at $T=\teff$. The relative differences between these curves are most 
   pronounced in the upper Hubbard band (UHB). Due to thermal fluctuations, the latter acquires 
   much more weight for $T=\teff$ than for the initial equilibrium state $T=0.2$. For the two
   pump-excited states, the weight of the UHB is comparable to the weight of the UHB at 
   $T=\teff$, but its distribution depends on the pumping process in such a way that the 
   spectral density at high energies is increased for the high-frequency pump. 
  
    \begin{figure}
  \centerline{\includegraphics[clip=true,width=0.99\columnwidth]{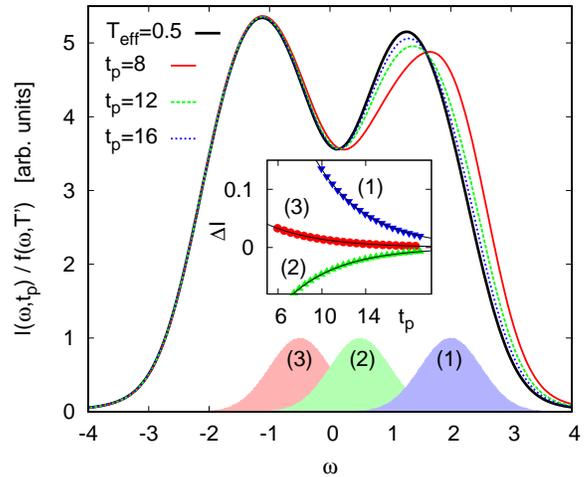}}
  \caption{Time-resolved photoemission spectrum [Gaussian probe, $\delta=2$]  
  of a pump-excited system ($U=2.5$, $\beta=5$), using the pump parameters of 
  Fig.~\ref{fig-doccrelax}a ($\Omega=5\pi/4$, $E_0$=1.9123). Main plot: $I(\omega,t_p)$ 
  for various probe-times $t_p$, compared to the spectrum $I_{\teff}(\omega)$
  of the equilibrium state at $\teff=0.5$. For better visibility of the upper Hubbard band,
  data have been divided by a Fermi-function with temperature $T'=0.55$. ($T'$ is chosen
  by hand and is slightly larger than $\teff$, in order to account for the broadening effect 
  due to the finite pulse duration.) Inset: Relative difference $I(\omega,t_p)$ to 
  $I_{\teff}(\omega)$ as a function of $t_p$, for three probe frequencies
  $\omega_{(1)}=2$, $\omega_{(2)}=0.5$, $\omega_{(3)}=-0.5$. Solid lines
  show exponential fits $I_{\Omega}(\omega)-I_{\teff}(\omega) \propto \exp(-t/\tterm)$,
  with parameters $\tterm$ = 4.70 ($\omega_{(1)}$),  4.82  ($\omega_{(2)}$),
  4.77 ($\omega_{(3)}$).}
  \label{fig-spec2}
  \end{figure}

   The differences between the curves in Fig.~\ref{fig-spec1} clearly confirm the nonthermal 
   nature of the excited states, which had already been deduced from the value of the double 
   occupancy. However, a thermal state at $T=\teff$ might not be accessible in a typical pump-probe 
   experiment, because electronic temperatures can be so high that heating the whole sample 
   to those temperatures would result in its destruction, or at least in a strong modification of its 
   properties. It is thus important to note that no quantitative comparison to the thermal state 
   at $T=\teff$ is required in order to identify a lack of thermalization of the electrons, but 
   it is enough to demonstrate that a state still has detailed memory on the way it was 
   prepared. In the present case this memory becomes evident from the fact that the spectral 
   weight distribution depends on the pump-frequency, for systems which have absorbed the 
   same amount of energy.
   
   At intermediate coupling, the time-evolution of the double occupancy has indicated a more 
   rapid thermalization of the system, and we will now demonstrate that this result is confirmed
   by the behavior of the spectral function. For this purpose we choose a state in the 
   crossover regime ($U=2.5$, $\beta=5$), excite it by a pulse of frequency $\Omega=U\pi/2$ 
   to the energy corresponding to $\teff=0.5$, and compare the spectrum $I(\omega,t_p)$ to the 
   spectrum $I_{\teff}(\omega)$ of the thermal equilibrium state at $T=\teff$ (Fig.~\ref{fig-spec2}). 
   At these parameters, the Mott gap is already closed, but the Hubbard bands are still
   separated by a dip in the spectral function. In the photoemission spectrum, the UHB appears 
   merely as a shoulder on the upper edge, but it can be enhanced by dividing the spectrum by a 
   Fermi function. For small probe-times, $I(\omega,t_p)$ differs from $I_{\teff}(\omega)$ by
   a shift of the UHB to larger frequencies, but in contrast to the results for $U=5$  we now 
   observe a relaxation of the spectrum towards the thermal equilibrium spectrum at later $t_p$. 
   In the inset of  Fig.~\ref{fig-spec2} we plot the relative difference $\Delta I = [I(\omega,t_p)-
   I_{\teff}(\omega)]/I_{\teff} (\omega)$ as a function of probe time $t_p$ for three frequencies, 
   $\omega_{(1)}=2$ (above the center of the UHB), $\omega_{(2)}=0.5$ (below the center of 
   the UHB), and $\omega_{(2)}=-0.5$ . Due to the shift of spectral weight from higher to lower 
   frequencies, $I(\omega_{(1)},t_p)$ decreases with $t_p$, while $I(\omega_{(2)},t_p)$ increases. A fit of the 
   curves with an exponential decay gives  relaxation times $\tterm= 4.70$ for $\omega_{(1)}$,  
   $\tterm=4.82$ for $\omega_{(2)}$, and $\tterm=4.77$ for $\omega_{(3)}$, comparable 
   to the relaxation time extracted from $d(t)$ for the same parameters, $\tterm$=4.87 
   (Fig.~\ref{fig-doccrelax}).

  \section{Conclusion}
  \label{sec::conclusion}

  In conclusion, we have investigated the relaxation of a Mott insulator subsequent to a strong laser excitation 
  over the Hubbard gap. After an initial strongly damped transient, the time-evolution of the double occupancy
  turns into an exponential relaxation towards its thermalized value at large times. For $U \gg W$, the thermalization 
  time grows rapidly with $U/W$, consistent with the exponentially small decay rate for one doublon in front of a  
  background of singly occupied sites and holes.\cite{Sensarma2010a} When $U$ is comparable to $W$, on the other 
  hand, the double occupancy thermalizes within a few times the inverse bandwidth. This more efficient thermalization 
  does not only become manifest through a decrease of the thermalization rate $1/\tterm$, but moreover, the system approaches 
  a point at which the exponential relaxation is entirely absent, and the thermal state is approached along a qualitatively 
  different, ``rapid'' pathway. 
  
  Exponential thermalization arises when the relevant kinetic equations 
  can be linearized around the thermal state, i.e., when relaxation close to a thermal equilibrium state is described 
  in terms of small deviations from that state. The fact that the long-time relaxation of the double occupancy can be 
  understood from the decay rate of single doublons fits into this scheme. On the other hand, the appearance of the rapid thermalization instead 
  of the exponential relaxation is quite remarkable, in particular as it appears at a value of $U$ that is far from any known 
  phase transition in equilibrium. It remains to be seen whether similar results can be obtained for weak coupling (which 
  is not accessible with the impurity solver used in this work), where the long-time behavior should be described by a 
  Boltzmann equation,\cite{Moeckel2008a} and whether it is related to the ``rapid thermalization'' at the dynamical 
  transition after an interaction quench in the Hubbard model. In any case, the analysis of the long-time tail 
  in terms of the ``residue'' $d_0$ used in this paper allows to rigorously define a  ``rapid thermalization'', and 
  thus should be useful for a future analysis of the dynamical transition after 
  the interaction quench in the Hubbard model and other models.
  
  In this work we have also demonstrated that the differences in thermalization behavior 
  are in principle  observable by means of time-resolved photoemission spectroscopy. 
  Quantitative differences between the time-dependent spectrum and equilibrium spectra are 
  small, but nonthermal behavior of  the pump-excited system may also be detected from 
  qualitative features, such as the memory of the electronic system on the precise form of 
  the excitation. We suppose that similar information can be drawn from optical 
  measurements, since both the conductivity and the photoemission spectrum 
  are derived from the same single-particle Green function (up to vertex corrections
  which can be controlled to some extent by varying the relative polarization of pump
  and probe).\cite{Eckstein2008b} Of course, most materials are more complicated than
  the simple one-band Hubbard model studied in this work. But our results suggest that 
  a similar pronounced dependence of the relaxation on  the interaction parameter occurs
  in more complex models. For example, it would be interesting to see whether the relaxation 
  of photo-doped metallic phases to insulating phases in charge transfer
  insulators\cite{Iwai2003a} is subject to a similar strong dependence  on the system 
  parameters as the thermalization of doublons in the half-filled single-band system.
  
  \section*{Acknowledgements}
  
  We thank M.~Kollar, C.~Kollath, T.~Oka, L.~Tarruell, and N.~Tsuji for useful discussions. 
  Numerical calculations were run on the Brutus cluster at ETH Zurich. We acknowledge 
  support from the Swiss National Science Foundation (Grant PP002-118866).

 \end{document}